\documentclass[12pt]{article}
\usepackage{cite}
\usepackage{amsmath}
  \numberwithin{equation}{section}
\usepackage{amssymb}
\usepackage{graphics}
\usepackage{bbold}
\usepackage{multirow}

\begin{document}

\begin{titlepage}

\begin{flushright}
TIT/HEP-633 \\
December 2013
\end{flushright}

\vspace{0.5cm}
\begin{center}
{\Large \bf
ODE/IM correspondence and modified affine Toda field equations
}
\lineskip .75em
\vskip2.5cm
{\large Katsushi Ito and Christopher Locke
}
\vskip 2.5em
{\normalsize\it Department of Physics,\\

Tokyo Institute of Technology\\
Tokyo, 152-8551, Japan} \vskip 1.0em
\vskip 3.0em
\end{center}

\begin{abstract}

We study the two-dimensional affine Toda field equations for affine Lie algebra $\hat{\mathfrak{g}}$ modified by a conformal transformation and the associated linear equations.
In the conformal limit, the associated linear problem reduces to a \mbox{(pseudo-)differential} equation.
For classical affine Lie algebra $\hat{\mathfrak{g}}$, we obtain a \mbox{(pseudo-)differential} equation corresponding to the Bethe equations for the Langlands dual of the Lie algebra $\mathfrak{g}$, which were found by Dorey et al. in study of the ODE/IM correspondence.

\end{abstract}

\end{titlepage}

\baselineskip=0.7cm
\section{Introduction}

The ODE/IM correspondence proposed by Dorey and Tateo \cite{Dorey:1998pt} gives an interesting relation between the spectral analysis of ordinary differential equations and the functional approach of two-dimensional quantum integrable models.
This correspondence is regarded as an example of the mysterious links between classical and quantum integrable models, which produce many interesting applications in supersymmetric gauge theories and superstring theories
  \cite{Gaiotto:2009hg, Nekrasov:2009rc, Teschner:2010je}.
For example, one can calculate gluon scattering amplitudes at strong coupling from the area of a null-polygonal minimal surface in AdS$_5$ by using the AdS/CFT correspondence, which is determined by nonlinear differential equations written in the form of the Hitchin system associated with SU(4).
The area is expressed in terms of solutions of the associated linear system and is evaluated through the Thermodynamic Bethe ansatz equations of the quantum integrable models characterized by a certain Y-system and Thermodynamic Bethe ansatz equations
  \cite{Alday:2009dv, Alday:2010vh} (for AdS$_3$ see also \cite{Hatsuda:2010cc}). 
In the case of lower dimensional AdS spaces, the minimal surface equation reduces to the modified sinh-Gordon equation for AdS$_3$
  \cite{Pohlmeyer:1975nb, DeVega:1992xc, Alday:2009yn}
and the $B_2$ affine Toda equation for AdS$_4$ \cite{Burrington:2009bh}.

In the ODE/IM correspondence, the spectral determinant of the differential equation leads to the Bethe ansatz equations in the conformal limit of certain integrable models \cite{Bazhanov:1994ft}.
The ordinary (pseudo-)differential operators for the integrable models related to classical simple Lie algebras of ABCD types were studied in \cite{Dorey:2006an}.
It was also pointed out that these (pseudo-)differential operators have a similar form to the Miura transformation of the scalar Lax operators in the Drinfeld-Sokolov reductions of affine Lie algebras \cite{Drinfeld:1984qv}.

Recently, Lukyanov and Zamolodchikov studied the spectral analysis of the classical sinh-Gordon equation modified by a conformal transformation and found that the linear system corresponds to the massive quantum sine-Gordon model \cite{Lukyanov:2010rn}.
This was generalized to a relation between the classical Tzitz\'eica-Bullough-Dodd equation and the quantum Izergin-Korepin model \cite{Dorey:2012bx}.

The sinh-Gordon and the Tzitz\'eica-Bullough-Dodd equations are the two-dimensional affine Toda field equations associated with affine Lie algebras $A_1^{(1)}$ and $A_2^{(2)}$ respectively.
The linear problems associated with the Lax representation of these affine Toda equations reduce to ordinary differential equations.
For the $A_1^{(1)}$ affine Toda equation, the corresponding ODE in the conformal limit is the Schr\"odinger equation with an angular momentum potential term \cite{Bazhanov:1998wj}.
For $A_2^{(2)}$, the ODE is the $A_2$-type third-order differential equation studied in \cite{Dorey:1999pv}.

In this paper, we will generalize the results of $A_1^{(1)}$ and $A_2^{(2)}$ affine Toda equations and study the link between classical affine Toda field equations associated with the affine Lie algebras which are modified by a conformal transformation, and their relation to (pseudo-)ordinary differential equations associated with the CFT limit of the integrable system.
We find that for the Langlands dual $\hat{\mathfrak{g}}^{\vee}$ of a classical affine Lie
algebra $\hat{\mathfrak{g}}$, the (pseudo-)ODE is shown to correspond to the one obtained by Dorey et al \cite{Dorey:2006an}.

This paper is organized as follows.
In section \ref{sec:affine_toda} we introduce the affine Toda field equations associated to affine Lie algebra $\hat{\mathfrak{g}}$.
In section \ref{sec:A} we study the simplest case of $A_r^{(1)}$ type algebra, deriving an ordinary differential equation from the linear problem, and show how in the conformal limit this leads to the Bethe ansatz equations.
The generalization to other classical affine Lie algebras is carried out in section \ref{sec:classical}.  Here the linear problem, and its associated (pseudo-)differential equation are derived.
Finally, in section \ref{sec:isomorphisms} the Lie algebra isomorphisms are shown to hold true in our formalism at both the level of the linear problem and the associated (pseudo-)differential equation.
In appendix \ref{appendix:affine_lie}, we summarize the matrix representations for classical affine Lie algebras, while in appendix \ref{appendix:exceptional}, we discuss the ODEs associated with the exceptional affine Lie algebra $G_2^{(1)}$ and its Langlands dual $D_4^{(3)}$.

\section{Affine Toda Field Equations} \label{sec:affine_toda}

We begin by fixing the notation used for Lie algebras in this paper (see \cite{kac1994infinite,fuchs-schweigert-2003}).
Let ${\mathfrak{g}}$ be a simple Lie algebra of rank $r$.
The generators  $\{E_{\alpha}, H^i\}$  ($\alpha\in \Delta$,  $i=1,\cdots r$) of ${\mathfrak{g}}$, where $\Delta$ is the set of roots, satisfy the commutation relations
\begin{align}
  & [ E_{\alpha}, E_{\beta} ]=N_{\alpha,\beta} E_{\alpha+\beta}, \quad \mbox{for $\alpha+\beta\neq 0$, } \\
  &  [ E_{\alpha}, E_{-\alpha} ]= \frac{2\alpha\cdot H}{\alpha^2},\\
  & [H^i, E_{\alpha}]=\alpha^i E_{\alpha}.
\end{align}
Here the structure constants $N_{\alpha,\beta}$ are only nonzero for $\alpha+\beta\in \Delta$.
Let $\alpha_1, \cdots, \alpha_r$ be the simple roots of $\mathfrak{g}$.
The Cartan matrix of $\mathfrak{g}$ is defined by
\begin{equation}
  A_{ij}={2\alpha_i\cdot \alpha_j\over \alpha_j^2}=\alpha_i\cdot \alpha_j^{\vee} \,,
\end{equation}
where $\alpha^{\vee}=\frac{2\alpha}{\alpha^2}$ is the coroot of $\alpha$.
We normalize the long root of $\mathfrak{g}$ to have length squared of 2.

The extended Dynkin diagram of $\mathfrak{g}$ is obtained by adding the root $\alpha_0 = -\theta$ where $\theta$ is the highest root, giving the affine Lie algebra $\hat{\mathfrak{g}}$.
The integers $n_i$ ($n_i^{\vee}$), called the (dual) Coxeter labels, are defined to satisfy
  $0 = \sum_{i=0}^{r} n_i\alpha_i = \sum_{i=0}^{r}n_i^{\vee}\alpha_i^{\vee}$.
In most cases $n_0 = 1$, but for the twisted algebra $A_{2r}^{(2)}$, $n_0^\vee = 2$.
The (dual) Coxeter number denoted as $h$ ($h^{\vee}$) is defined by
\begin{equation}
  h=\sum_{i=0}^{r}n_i, \quad h^{\vee}=\sum_{i=0}^{r} n_i^{\vee} .
\end{equation}
The (co)fundamental weights $\omega_i$ ($\omega^{\vee}_i$) are the vectors dual to $\alpha^{\vee}_i$ ($\alpha_i$) satisfying
\begin{align}
  \omega_i\cdot\alpha^{\vee}_j=\delta_{ij}, \quad \omega^{\vee}_i\cdot\alpha_j=\delta_{ij}.
\end{align}
We define the (co)Weyl vector $\rho$ ($\rho^{\vee}$) as the sum of the (co)fundamental weights.

\subsection{Modified affine Toda equation} \label{subsec:mToda}

The Lagrangian of the affine Toda field theory associated to $\hat{\mathfrak{g}}$ is \cite{Mikhailov:1981,Wilson:1981,OliveTurok:1983}
\begin{equation}
  {\cal L} = {1\over2} \partial^\mu\phi \cdot \partial_\mu\phi - \left({m\over\beta}\right)^2 \sum_{i=0}^{r}n_i \left[\exp(\beta \alpha_i \cdot \phi)-1\right] , \label{eq:Lagrangian}
\end{equation}
where $\phi$ is an $r$-component scalar field, $m$ is a mass parameter and $\beta$ a dimensionless coupling parameter.
The equation of motion is
\begin{equation}
  \partial^\mu \partial_\mu \phi + \left( {m^2\over\beta} \right) \sum_{i=0}^{r} n_i \alpha_i \exp(\beta \alpha_i\phi) = 0 \,. \label{eq:todaeq1}
\end{equation}
Expanding the field as $\phi = \sum_{i=1}^{r}\alpha_i^\vee \phi_i$, it can be found that the component fields $\phi_i = \omega_i \cdot \phi$ ($i = 1,\ldots, r$) have equation of motion
\begin{equation}
  \partial^\mu \partial_\mu \phi_i + \left( {m^2\over\beta} \right) n_i^\vee \exp\left(\beta \sum_{j=1}^{r}A_{ij}\phi_j\right) - \left( {m^2\over\beta} \right) n_i^\vee \exp\left(-\beta \sum_{i,j=1}^{r} n_i A_{ij} \phi_j \right) = 0 \,.
\label{eq:todaeq2}
\end{equation}

From this point on, we will use complexified coordinates $(z,\bar{z})$ given by
\begin{equation}
  z = {1\over2}(x^0+i x^1), \quad \bar{z} = {1\over2}(x^0-i x^1).
\end{equation}
Note that this gives $\partial^\mu \partial_\mu = \partial \bar\partial$.

Toda field theories associated with simple Lie algebras are massless and possess conformal symmetry.
However, for affine Toda field theories the conformal symmetry is broken and the theories are massive.
Using a conformal transformation, the affine Toda field equation can be put into a modified form which has been used in recent papers
  \cite{Lukyanov:2010rn, Dorey:2012bx,Lukyanov:2013wra}.
This form is obtained by doing a conformal transformation along with a field redefinition,
\begin{gather}
  z \rightarrow \tilde{z} = f(z) \,, \quad \phi \rightarrow \tilde{\phi} = \phi - {1\over\beta} \rho^\vee \log(\partial f \bar{\partial}\bar{f}) \,,
\end{gather}
where $\rho^\vee$ is the co-Weyl vector with the properties
\begin{gather}
  \rho^\vee \cdot \alpha_i = 1 \,, \quad\quad \rho^\vee \cdot \alpha_0 = -\sum_{i=1}^{r} n_i = 1 - h \,.
\end{gather}
Using the above conformal transformation, we get the modified affine Toda field equation
\begin{gather}
  \partial\bar{\partial}\phi + \left({m^2\over\beta}\right) \left[\sum_{i=1}^{r} n_i \alpha_i \exp(\beta \alpha_i\phi) + p(z)\bar{p}(\bar{z}) n_0 \alpha_0 \exp(\beta \alpha_0\phi) \right] = 0 , \label{eq:mtodaeq1}
\end{gather}
where the conformal factors have been absorbed into the definitions of $p(z)$ and $\bar{p}(z)$,
\begin{gather}
  p(z) = (\partial f)^h, \quad \bar{p}(\bar{z}) = (\bar{\partial}\bar{f})^h .
\end{gather}

The equation of motion can be written as a zero curvature condition
  $F = \mathrm{d} \mathbf{A} + \mathbf{A} \wedge \mathbf{A} = 0$
  for the one-form
  $\mathbf{A} = A \, \mathrm{d}z + \bar{A} \, \mathrm{d} \bar{z}$.
Our choice of connection that gives this modified affine Toda equation \eqref{eq:mtodaeq1} is
\begin{gather}
  A = {\beta\over 2}\partial \phi\cdot H + m e^{\lambda} \left\{ \sum_{i=1}^{r} \sqrt{n_i^\vee} E_{\alpha_i} \exp\left({\beta\over 2}\alpha_i\phi\right) + p(z) \sqrt{n_0^\vee} E_{\alpha_0} \exp\left({\beta\over 2}\alpha_0\phi\right) \right\} \label{eq:connection_A} ,\\
  \bar{A} = -{\beta\over 2}\bar{\partial} \phi\cdot H - m e^{-\lambda} \left\{\sum_{i=1}^{r} \sqrt{n_i^\vee} E_{-\alpha_i} \exp\left({\beta\over 2}\alpha_i\phi\right) + \bar{p}(\bar{z}) \sqrt{n_0^\vee} E_{-\alpha_0} \exp\left({\beta\over 2}\alpha_0\phi\right) \right\}.\nonumber\\
\end{gather}
Here we have introduced the spectral parameter $e^{\lambda}$.

We are interested in the linear problem $(\mathrm{d} + \mathbf{A}) \Psi = 0$.
When we later take explicit matrix representations for the generators $H$ and $E_\alpha$, we will write out this linear problem as two equations coming from the terms proportional to $\mathrm{d}z$ and $\mathrm{d}\bar{z}$: $(\partial + A) \Psi = 0$ and $(\bar\partial + \bar{A}) \Psi = 0$.

\subsection{Asymptotic behavior of modified affine Toda equation} \label{subsec:mToda_asym}

We will let the function $p(z)$ have the form
\begin{equation}
\begin{gathered}
  p(z) = z^{hM}-s^{hM}, \quad \bar{p}(\bar{z}) = \bar{z}^{hM} - \bar{s}^{hM} \label{eq:functionp1}  
\end{gathered}
\end{equation}
for some positive real number $M > \tfrac{1}{h-1}$ and parameter $s$.
This is a generalization of the $A_1^{(1)}$ case studied in \cite{Lukyanov:2010rn}, and $A_2^{(2)}$ case of \cite{Dorey:2012bx}.

Using the above form of $p(z)$ motivates us to define the following transformation,
\begin{align}
  \hat{\Omega}_k : \; \left\{
  \begin{matrix}
    & z \rightarrow z \ e^{2\pi k i / hM} \\
    & s \rightarrow s \ e^{2\pi k i / hM} \\
    & \lambda \rightarrow \lambda - \tfrac{2\pi k i}{hM}
  \end{matrix} \right. \;. \label{eq:mToda_sym}
\end{align}
$\hat{\Omega}_k$ leaves the equation of motion and linear problem unchanged for integer $k$, so the modified affine Toda linear problem is symmetric under this transformation.

We assume that the large $|z|$ asymptotic solution of the equation of motion \eqref{eq:mtodaeq1} is
\begin{gather}
  \phi(z,\bar{z}) = \frac{M \hat{\rho}}{\beta} \log(z\bar{z}) + \mathcal{O}(1) .
\end{gather}
For $z$ approaching $0$, we expand the field $\phi$ as
\begin{gather}
  \phi(z,\bar{z}) = g \log(z\bar{z}) + \phi^{(0)}(g) + \gamma(z,\bar{z},g) + \sum_{i=0}^r \frac{C_i(g)}{(c_i(g) + 1)^2} (z\bar{z})^{c_i(g)+1} + \cdots , \label{eq:mtoda_small}
\end{gather}
where the extra dots represent terms that are higher order in $z\bar{z}$ than those explicitly in the sum.
$g$ is a vector that controls the asymptotic behavior near $0$, and $\gamma$ is defined in such a way as to protect the symmetry under \eqref{eq:mToda_sym} for integer $k$, and under $z \leftrightarrow \bar{z}$,
\begin{gather}
  \gamma(z,\bar{z},g) = \sum_{i=1}^\infty \gamma_k(g) (z^{hMk} + \bar{z}^{hMk}) .
\end{gather}
Substituting this limiting behavior \eqref{eq:mtoda_small} into the modified affine Toda equation \eqref{eq:mtodaeq1} gives
\begin{equation}
\begin{gathered}
  \sum_{i=0}^r C_i (z \bar{z})^{c_i} + \frac{m^2}{\beta} \sum_{i=1}^r n_i \alpha_i (z\bar{z})^{\beta\alpha_i \cdot g} e^{\beta \alpha_i \cdot \phi^{(0)} + \cdots} \\
  - \frac{m^2}{\beta} (s\bar{s})^{hM} n_0 \alpha_0 (z\bar{z})^{\beta\alpha_0 \cdot g} e^{\beta \alpha_0 \cdot \phi^{(0)} + \cdots} + \cdots = 0 . \label{eq:mtodaeq2}
\end{gathered}
\end{equation}
The $c_i$ terms must match the exponents $\beta\alpha_i \cdot g$ to cancel out, so $c_i = \beta\alpha_i \cdot g$ for $i = 0, 1, \ldots, r$, with the constants $C_i$ forced to be
  $C_i = -\tfrac{m^2}{\beta} n_i \alpha_i e^{\beta \alpha_i \cdot \phi^{(0)}(g)}$
  for $i = 1, 2, \ldots, r$, and
  $C_0 = \tfrac{m^2}{\beta} (s\bar{s})^{hM} n_0 \alpha_0 e^{\beta \alpha_0 \cdot \phi^{(0)}(g)}$.
Also, since the asymptotic behavior is assumed to be logarithmic at leading order, the exponents $c_i+1$ must be positive.
This means that allowed values of $g$ must obey the conditions
\begin{gather}
  \beta\alpha_i \cdot g + 1 > 0 \,, \quad (i = 0, 1, \ldots, r). \label{eq:small_restriction}
\end{gather}

\section{$A_r^{(1)}$ modified affine Toda equation} \label{sec:A}

Now we will study the solutions of the linear equations $(\partial + A) \Psi = 0$ and $(\bar\partial + \bar{A}) \Psi = 0$ for the affine Lie algebras.
We will first consider in detail the simplest affine Lie algebra, $A_r^{(1)}$.
This will serve to introduce the methodology that will be used for the other algebras in subsequent sections.

\subsection{Linear equation} \label{sec:A_lin_eq}

$A_r^{(1)}$ is a simply laced affine Lie algebra with Coxeter number $h = r+1$.
We will explicitly write out the linear equations for the fundamental $(r+1)$-dimensional matrix representation of $A_r^{(1)}$, given in appendix \ref{appendix:affine_lie}.
The weight vectors $h_1,\cdots, h_{r+1}$ of the fundamental representation with the highest weight $h_1 = \omega_1$ satisfy
\begin{align}
  h_i-h_{i+1}=\alpha_i.
\end{align}

The connection can be simplified through the gauge transformation
\begin{gather}
  \tilde{A} = UA U^{-1} + U\partial U^{-1}, \quad \tilde{\Psi} = U \Psi , \label{eq:gauge}\\
  \quad U = diag( e^{-\tfrac{\beta}{2} h_1 \cdot \phi},\cdots, e^{-\tfrac{\beta}{2} h_{r+1} \cdot \phi}) .
\end{gather}
When this is done, the holomorphic connection becomes
\begin{align}
\tilde{A}=\left(
\begin{array}{ccccc}
{\beta}h_1 \cdot \partial\phi & me^{\lambda}
& 0 & \cdots&0 \\
0 &{\beta}h_2 \cdot \partial\phi & me^{\lambda}
 &  &\vdots\\
& & \ddots & & \\
\vdots & & & {\beta}h_r \cdot \partial\phi &me^{\lambda}
 \\
me^{\lambda}  p(z)&
 & &0 & {\beta}h_{r+1} \cdot \partial\phi\\
\end{array}
\right) .
\end{align}
To write out the component equations, it is useful to define a differential operator,
\begin{gather}
  D(a) \equiv \partial + \beta a \cdot \partial\phi .
\end{gather}
Using this notation, the component equations become
\begin{subequations}
\begin{gather}
  D(h_1) \tilde{\psi}_1 = -m e^{\lambda} \tilde{\psi}_{2} \label{eq:A_lin_comp_first} , \\
  \vdots\nonumber\\
  D(h_r) \tilde{\psi}_r = -m e^{\lambda} \tilde{\psi}_{r+1} , \label{eq:A_lin_comp_nextToLast} \\
  D(h_{r+1}) \tilde{\psi}_{r+1} = -m e^{\lambda} p(z) \tilde{\psi}_1 . \label{eq:A_lin_comp_last}
\end{gather}
\end{subequations}
The linear equations for $\tilde{\Psi}$ can be combined into a single differential equation on just $\tilde{\psi}_1$.
\begin{gather}
  D(h_{r+1}) \cdots D(h_{1}) \tilde{\psi}_{1} = (-m e^{\lambda})^h p(z) \tilde{\psi}_{1} .
\end{gather}
Note that the differential operator on the left hand side is nothing but the scalar Lax operator of Gel'fand-Dickii for the generalized KdV equations \cite{Drinfeld:1984qv}.

A similar analysis can be done on the barred linear equation $(\bar\partial + \bar{A}) \Psi = 0$ after a gauge transformation given by
  $U = diag( e^{\tfrac{\beta}{2} h_1 \cdot \phi},\cdots, e^{\tfrac{\beta}{2} h_{r+1} \cdot \phi})$.
In this case, the compatibility condition becomes
\begin{gather}
  \bar{D}(-h_1) \cdots \bar{D}(-h_{r+1}) \tilde{\bar\psi}_{r+1} = (m e^{-\lambda})^h \bar{p}(\bar{z}) \tilde{\bar\psi}_{r+1} .
\end{gather}
To simplify notation, we define $\psi = \tilde{\psi}_1$, and $\bar{\psi} = \tilde{\bar{\psi}}_{r+1}$.
The resulting compatibility conditions are then
\begin{align}
  D(h_{r+1}) \cdots D(h_1) \psi &= (-m e^{\lambda})^h p(z) \psi \label{eq:A_comp_eq2} \\
  \bar{D}(-h_1) \cdots \bar{D}(-h_{r+1}) \bar\psi &= (m e^{-\lambda})^h \bar{p}(\bar{z}) \bar\psi \label{eq:A_compbar_eq2} .
\end{align}
Note the that these two differential operators are adjoint of each other.

\subsection{Asymptotic behavior of the solutions} \label{subsec:A_asym}

We examine the asymptotic behavior of the differential equation (\ref{eq:A_comp_eq2}) for large and small $z$.
Substituting the small-$z$ asymptotic behavior of $\phi$ into the equation \eqref{eq:A_comp_eq2} gives, to leading order,
\begin{gather}
  \left(\partial + \tfrac{\beta}{z} h_{r+1} \cdot g \right) \cdots \left(\partial + \tfrac{\beta}{z} h_{1} \cdot g \right) \psi = 0 \,.
\end{gather}
An indicial equation can then be found by substituting the ansatz $\psi \sim z^\mu$, with roots given by
\begin{gather}
  \mu_i = i - \beta h_{i+1} \cdot g \quad \text{for} \quad i = 0, 1, \ldots, r \,.
\end{gather}
These roots can be shown to be well ordered using equation \eqref{eq:small_restriction},
\begin{gather}
  \mu_{i+1} - \mu_i = 1 - \beta (h_{i+2} - h_{i+1}) \cdot \nu = 1 + \beta \alpha_{i+1} \cdot g > 0, \quad \mbox{$(i=0,\ldots,r-1)$} \label{eq:indicial_mu} .
\end{gather}
Similarly, the asymptotic behavior of $\bar\psi \sim \bar{z}^{\bar\mu}$ has possible values $\bar{\mu}_i = i + \beta h_{r+1-i} \cdot g$ for $i = 0, 1, \ldots, r$, which are well ordered as well.

In the large $z$ limit, only the highest derivative term of equation \eqref{eq:A_comp_eq2} and the term proportional to $p(z)$ survive.
The resulting differential equation is
\begin{gather}
  (\partial^{r+1} + (-1)^{r} (m e^\lambda)^{h} p(z)) \psi = 0,
\end{gather}
which was obtained in \cite{Suzuki:1999hu,Dorey:2000ma}.
A WKB analysis gives the large $z$ asymptotic behavior
\begin{align}
  \psi & \sim z^{-\frac{rM}{2}} \exp\left( -\frac{z^{M+1}}{M+1} me^\lambda + g(\bar{z}) \right) ,
\end{align}
where $g(\bar{z})$ is an anti-holomorphic function.
This solution is valid in the Stokes sector
\begin{gather}
  |\arg z| < \frac{(r+2)\pi}{(r+1)(M+1)}. \label{eq:asym_decay_sector}
\end{gather}
An analogous WKB analysis for the anti-holomorphic equation \eqref{eq:A_compbar_eq2} gives
\begin{align}
  \bar\psi \sim \bar{z}^{-\frac{rM}{2}} \exp\left( -\frac{\bar{z}^{M+1}}{M+1} me^{-\lambda} + f(z) \right) .
\end{align}

Equation \eqref{eq:A_lin_comp_last} can be used to connect the two asymptotics of $\psi_1$ and $\psi_{r+1}$.
This fixes the functions $f$ and $g$ and shows that the constants of proportionality agree.
The result, after using the definitions of $\tilde{\psi}_i$ and $\tilde{\bar{\psi}}_i$, the fact that $\rho^\vee \cdot h_i = \tfrac{r - 2(i-1)}{2}$, is
\begin{gather}
  \psi_i = \left(\frac{z}{\bar{z}}\right)^{-\tfrac{(r-2(i-1))M}{4}} \exp\left( -\frac{z^{M+1}}{M+1} me^{\lambda} - \frac{\bar{z}^{M+1}}{M+1} me^{-\lambda}\right) .
\end{gather}
This gives a unique asymptotically decaying solution in Stokes sector \eqref{eq:asym_decay_sector} with large $|z|$ behavior given by (using polar coordinates $z = \rho e^{i\theta}$),
\begin{gather}
  \Xi(\rho,\theta | \lambda) \sim C \begin{pmatrix} e^{-\tfrac{irM\theta}{4}}\\
  e^{-\tfrac{i(r-2)M\theta}{4}}\\
  \vdots\\
  e^{\tfrac{irM\theta}{4}}\end{pmatrix}
  \exp\left( -\frac{2\rho^{M+1}}{M+1} \ m \cosh(\lambda  + i\theta(M+1)) \right) \label{eq:Xi}.
\end{gather}

Next, using the small $z$ asymptotics $\psi \sim z^{\mu_i}$ allows one to define a basis of vectors for solutions to the linear equations in which $\Xi$ will be expanded.
As a first step, consider taking $\tilde{\psi}_1 \sim z^{\mu_0} + \mathcal{O}(z^{\mu_0 + h})$ corresponding to the lowest indicial root $\mu_0$. Using the equations on $\tilde{\Psi}$ shows that $\tilde{\psi}_2 \sim 0$ to first order.
Converting this solution back to the pre-gauge transformed vector $\Psi$ gives $\psi_1 \sim (\bar{z} / z)^{\frac{\beta}{2} h_1 \cdot g}$.
Similarly, carrying out this argument for general indicial root $\mu_i$ gives a vector $\Psi$ with non-zero component $\psi_{i+1} \sim (\bar{z} / z)^{\frac{\beta}{2} h_{i+1} \cdot g}$.
This is the dominant component of the vector $\Psi$ at small $z$, because components of $\Psi$ for indices smaller than $(i+1)$ are higher powers of $z$ while successive components have their first order term eliminated due to the action of $D(h_{i+1})$ on $\tilde{\psi}_{r+1}$.  So to first order the dominant term of $\Psi$ is the $(i+1)$th component and we can set all other components to zero to this level of approximation.

Note that the above argument only considers the holomorphic equations.
However a comparison of $\psi_{r+1} \sim (\bar{z} / z)^{\frac{\beta}{2} h_{r+1} \cdot g}$ to $\bar{\psi} \sim \bar{z}^{\bar{\mu}_0}$ shows that the above form also has correct $\bar{z}$ behavior.
So the basis of solutions for small $z$, each denoted by $i = 1, \ldots r+1$, with asymptotic behavior determined by $\tilde{\psi}_1 \sim z^{\mu_{i-1}} + \mathcal{O}(z^{\mu_{i-1} + h})$, to dominant order is
\begin{gather}
  (\Psi^{(i)})_j \sim \delta_{ij} (\bar{z} / z)^{\frac{\beta}{2} h_{i} \cdot g} .
\end{gather}

Since these functions form a basis in the space of solutions for the linear problem, the unique decaying solution $\Xi$ can be expanded as
\begin{gather}
  \Xi = \sum_{i=0}^r Q_i(\lambda) \Psi^{(i)} .
\end{gather}
As per \cite{Lukyanov:2010rn,Dorey:2012bx}, we expect that these $Q$-coefficients should coincide with the the $Q$-functions of a 2D massive quantum field theory related to the Lie algebra $A_r$.  However, we will not pursue this investigation of the massive ODE/IM correspondence implied here, but will instead next go on to discuss the conformal limit and see its role in connecting with the differential equations of \cite{Suzuki:1999hu,Dorey:2000ma,Dorey:2006an}

\subsection{Conformal limit and Bethe ansatz} \label{sec:A_conf}

Recall that 
the modified affine Toda linear problem is symmetric under the
transformation  $\hat{\Omega}_k$ in \eqref{eq:mToda_sym} for integer $k$.
Using this 
we define a $k$-Symanzik rotated solution, where $k$ will be free to take any real value, to be
\begin{gather}
  \Psi_k(\rho, \theta | \lambda) = \hat{\Omega}_k \Psi(\rho, \theta | \lambda) \label{eq:Symanzik},
\end{gather}

Next, a series of auxiliary functions can be defined using the asymptotically decaying solution $\Xi$ from equation \eqref{eq:Xi},
\begin{gather}
  \psi^{(a)}_k = \mathrm{Det}_a [\Xi_{k+(1-a)/2}, \Xi_{k+(3-a)/2}, \ldots \Xi_{k+(a-1)/2}] .
\end{gather}
Here, the subscript $\mathrm{Det}_a$ shows that we only take the upper $a$ components of each vector, so that the matrix is square,
while the subscript on $\Xi$ denotes a Symanzik rotation, as per equation \eqref{eq:Symanzik}.
For $A_r^{(1)}$-type Lie algebra, it can be shown that $\psi^{(r+1)}$ as defined above is constant.
This follows because $\partial \psi_{r+1} \sim \psi_1$, so the derivative of the determinant $\psi^{(r+1)}$ is zero.
An appropriate renormalization then gives $\psi^{(r+1)} = 1$.

Using some properties of determinants (see \cite{Dorey:2006an}), it can be shown that these auxiliary functions satisfy the relations
\begin{gather}
  \psi^{(a-1)} \psi^{(a+1)} = W^{(2)}[ \psi^{(a)}_{-1/2}, \psi^{(a)}_{1/2} ] \label{eq:A_aux_eq} ,
\end{gather}
where as noted $\psi^{(r+1)} = 1$, $W^{(2)}[f,g]$ is just the Wronskian, and for notational convenience we set $\psi^{(0)} = 1$.

The conformal limit is defined through the definitions
\begin{equation}
\begin{gathered}
  x = (me^{\lambda})^{1/(M+1)} z, \;\; E = s^{hM} (me^\lambda)^{hM/(M+1)} \,,\\
  \tilde{x} = (me^{-\lambda})^{1/(M+1)} \bar{z}, \;\; \tilde{E} = s^{hM} (me^{-\lambda})^{hM/(M+1)} \,.
\end{gathered} \label{eq:conf_trans}
\end{equation}
This transformation gives 
  $D(a) = (me^{\lambda})^{1/(M+1)}(\partial_x + \beta a \cdot \partial_x\phi)$,
  and for $p(x,E) \equiv x^{hM} - E$ we have
  $p(z,s) = (m e^\lambda)^{\tfrac{-hM}{M+1}} p(x,E)$.
Then, the light-cone limit $\bar{z} \rightarrow 0$ is taken while sending $\lambda \rightarrow \infty$.
Afterwards $z$ is taken to $0$ while keeping $x$ and $E$ finite.
In this limit, using the small-$\rho$ behavior of $\phi$ gives $D_x(a) = \partial_x + \beta \tfrac{a \cdot g}{x}$, and the compatibility equation becomes:
\begin{align}
  \left[ D_x(h_{r+1}) \cdots D_x(h_1) - (-1)^h p(x,E) \right] y(x,E,g) = 0 \label{eq:MToda_conf_eq} .
\end{align}
This is the same differential equation studied in \cite{Suzuki:1999hu,Dorey:2000ma}, and also in \cite{Dorey:2006an}.
This equation is the ODE part of the ODE/IM correspondence in the case of $A_r$ type Lie algebras.
A WKB analysis shows that this has a unique subdominant solution valid in Stokes sector \eqref{eq:asym_decay_sector} of
\begin{gather}
  y(x,E,g) \sim C x^{-\tfrac{rM}{2}} \exp\left( -\tfrac{x^{M+1}}{M+1} \right) \label{eq:conf_y} ,
\end{gather}

Noting that the small $x$ behavior of equation \eqref{eq:conf_y} has indicial exponents given by equation \eqref{eq:indicial_mu}, a basis of solutions to this equation can be defined
\begin{gather}
  \chi^{(i)} \sim x^{\mu_i} + \mathcal{O}(x^{\mu_i + h}) , \quad\quad i = 0, \ldots r \,.
\end{gather}
At this point, one can follow the argument of \cite{Dorey:2006an} to arrive at the Bethe ansatz equations
\begin{gather}
  \omega^{\mu_{i-1} - \mu_i}
  \frac{Q^{(i-1)}_{-1/2}(E^{(i)}_n) Q^{(i)}_{1}(E^{(i)}_n) Q^{(i+1)}_{-1/2}(E^{(i)}_n)}
  {Q^{(i-1)}_{1/2}(E^{(i)}_n) Q^{(i)}_{-1}(E^{(i)}_n) Q^{(i+1)}_{1/2}(E^{(i)}_n)}
  = -1 \,.
\end{gather}
Here the lower indices on $Q^{(i)}$ denote a Symanzik rotation, $\omega =e^{2\pi i / h(M+1)}$, and $E_n^{(i)}$ are the set of zeros for $Q^{(i)}$.
To write the exponent of $\omega$ using the Cartan matrix, we make the identification
\begin{gather}
  \gamma \equiv-\rho^\vee - \beta g \label{eq:A_gamma_def} ,
\end{gather}
which, using the expansion $\gamma = \alpha_j^\vee \gamma_j$, implies $\mu_{i-1} - \mu_i = A_{ij} \gamma_j$.
Recalling that $\mu_i = i - \beta h_{i+1} \cdot g$, the following equation is consistent with this definition
\begin{gather}
  \beta \omega_i \cdot g = \sum_{j=0}^{i-1} (j - \mu_j) \,.
\end{gather}
Substituting this into the definition of $\gamma$ gives, in agreement with \cite{Dorey:2006an},
\begin{gather}
  \gamma_i = \omega_i \cdot \gamma = \sum_{j=0}^{i-1} \mu_j - \frac{j(h-1)}{2} \,.
\end{gather}
The result is the Bethe ansatz equations in the form
\begin{gather}
  \prod_{j=1}^r \omega^{A_{ij} \gamma_j} \frac{Q^{(j)}_{A_{ij} / 2} (E_n^{(i)})} {Q^{(j)}_{-A_{ij} / 2} (E_n^{(i)})} = -1 . \label{eq:A_BAE}
\end{gather}

This section has shown how starting with the linear problem associated with the $A_r^{(1)}$ type affine Toda theory modified by a conformal transformation, one can arrive at a specific ordinary differential equation.
By taking the conformal limit of this ODE and using some special functional relations, the Bethe ansatz equations can be derived.
This connection between the spectral ODE problem and Bethe ansatz equations is an example of the ODE/IM correspondence.

\section{Linear problem for classical affine Lie algebras} \label{sec:classical}

In this section, we discuss the linear problem for other classical affine Lie algebras.
Explicit fundamental matrix representations and Coxeter labels are summarized in appendix \ref{appendix:affine_lie}.

\subsection{$D_r^{(1)}$ modified affine Toda equation} \label{sec:D}

The affine Lie algebra $D_r^{(1)}$ is simply laced and has Coxeter number $h = 2r-2$.
The fundamental $2r$-dimensional matrix representation of $D_r$ is given in appendix \ref{appendix:affine_lie}.
For the representation with highest weight $h_1 = \omega_1$, the weight vectors $h_{1},\cdots, h_{2r}$ satisfy
\begin{equation}
\begin{aligned}
  & h_i-h_{i+1} = \alpha_i \quad (i = 1, \cdots, r-1), \\
  & h_{2r+1-i} = -h_i, \quad (i = 1, \cdots, r).
\end{aligned}
\end{equation}

After doing gauge a transformation \eqref{eq:gauge} with
\begin{gather}  
  U = diag( e^{-\tfrac{\beta}{2} h_1 \cdot \phi},\cdots, e^{-\tfrac{\beta}{2} h_{r} \cdot \phi}, e^{\tfrac{\beta}{2} h_r \cdot \phi},\cdots, e^{\tfrac{\beta}{2} h_{1} \cdot \phi}),
\end{gather}
the linear equation $(\partial + \tilde{A}) \tilde{\Psi} = 0$ gives a set of differential equations.
To write out an equation on just $\tilde{\psi}_1$, it is useful to make the following identifications,
\begin{gather}
  \mathbf{h} = (h_r, \ldots, h_1) \,, \quad \mathbf{h}^\dagger = (-h_1, \ldots, -h_r) \,,\\
  D(\mathbf{h}) = D(h_r) \cdots D(h_1) .
\end{gather}
Writing out $D(\mathbf{h}^\dagger)$ shows that it is, within a sign factor, the adjoint  $D(\mathbf{h})^\dagger$.
Using this notation, the following relations can then be shown to follow from the system of differential equations,
\begin{gather}
  D(\mathbf{h}) \tilde{\psi}_1 = 2 (-m e^\lambda)^{r-1} \left( \prod_{i=1}^r \sqrt{n_i} \right) \partial \tilde{\psi}_{r+1} ,\\
  D(\mathbf{h}^\dagger) \tilde{\psi}_{r+1} = 2 (-m e^\lambda)^{r-1} \left( \prod_{i=1}^r \sqrt{n_i} \right) \sqrt{p(z)} \partial \sqrt{p(z)} \tilde{\psi}_{1} .
\end{gather}
Defining $\psi = \tilde{\psi}_1$, an equation on $\psi$ can be written,
\begin{gather}
  D(\mathbf{h}^\dagger) \partial^{-1} D(\mathbf{h}) \psi = 2^{r-1} (m e^\lambda)^h \sqrt{p(z)} \partial \sqrt{p(z)} \psi \label{eq:D_compat_eq} .
\end{gather}
This pseudo-differential equation is in agreement with that found in \cite{Dorey:2006an} for $D$-type Lie algebras.

Substituting the small-$z$ asymptotic behavior of $\phi$ into equation \eqref{eq:D_compat_eq},
an indicial equation can be found by substituting the ansatz $\psi_1 \sim z^\mu$.
This equation has roots given by ($i=0,\ldots, r-1$)
\begin{gather}
  \mu_i = i - \beta h_{i+1} \cdot g ,\\
  \mu_{r+i} = i + r-1 + \beta h_{r-i} \cdot g .
\end{gather}
Following the argument of section \ref{subsec:A_asym} shows that $\mu_{i+1} > \mu_i$ and $\mu_{r+i+1} > \mu_{r+i}$.
However, the same thing does not happen for roots $\mu_{r}$ and $\mu_{r-1}$.
Calculating the difference gives
\begin{gather}
  \mu_r - \mu_{r-1} = 2 \beta h_r \cdot g = 2 \beta (\omega_r - \omega_{r-1}) \cdot g .
\end{gather}
It looks like the additional requirement of $h_r \cdot g > 0$ is necessary to guarantee that the roots of the indicial equation are well ordered.
In \cite{Dorey:2006an}, the constraint $\mu_i < h / 2$ for all $i = 0, \ldots, r-1$ was imposed.
In particular, for $i = r-1$, this constraint becomes $\mu_{r-1} = r-1 - \beta h_r \cdot g < r - 1$, or $h_r \cdot g > 0$.
However, considering the $\mu_r$ case instead gives $h_r \cdot g < 0$.
So for consistency one actually needs $h_r \cdot g = 0$ and $\mu_{r-1} = \mu_r$.
This only gives $2r-1$ unique roots to this equation, which is in agreement with the order of the pseudo-differential equation \eqref{eq:D_compat_eq} in the large $z$ limit, which is $2r-1$.

It turns out that the exact same transformation, \eqref{eq:conf_trans}, as the $A_r^{(1)}$ case allow the conformal limit to be taken.
When this is done, the compatibility equation becomes
\begin{gather}
  D_x(\mathbf{h}^\dagger) \partial_x^{-1} D_x(\mathbf{h}) \psi_1 = 2^{r-1} \sqrt{p(x,E)} \partial_x \sqrt{p(x,E)} \psi_{1} \label{eq:D_compat_x_eq} .
\end{gather}
The most subdominant solution to this equation can be denoted, like in section \ref{sec:A}, $y(x,E,g)$, which has in the appropriate Stokes sector the asymptotic behavior
\begin{gather}
  y(x,E,g) \sim C x^{-hM/2} \exp\left( -\frac{x^{M+1}}{M+1} \right) .
\end{gather}
Following the steps of \cite{Dorey:2006an} gives the same form of Bethe ansatz equations \eqref{eq:A_BAE} with $D$-type Cartan matrix, and $\gamma$ as defined in \eqref{eq:A_gamma_def}.

\subsection{$B_r^{(1)}$ modified affine Toda equation}

For affine Lie algebra of type $B_r^{(1)}$, we consider the $(2r+1)$-dimensional vector representation with highest weight $h_1=\omega_1$.
The weight vectors $h_1,\cdots, h_{2r+1}$ are then
\begin{equation}
\begin{aligned}
  & h_i-h_{i+1}=\alpha_i, \quad (i=1,\cdots, r-1) \\
  & h_{r+1}=0, \quad h_{2r+2-j}=-h_j, \quad (j=1,\cdots, r)
\end{aligned}
\end{equation}

After carrying out a gauge transformation \eqref{eq:gauge} with
\begin{gather}  
  U = diag( e^{-\tfrac{\beta}{2} h_1 \cdot \phi},\cdots, e^{-\tfrac{\beta}{2} h_{r} \cdot \phi}, 1, e^{\tfrac{\beta}{2} h_r \cdot \phi},\cdots, e^{\tfrac{\beta}{2} h_{1} \cdot \phi}),
\end{gather}
the linear equation $(\partial + \tilde{A}) \tilde{\Psi} = 0$ gives a set of differential equations.
These differential equations can be combined into a compatibility condition on $\psi = \tilde{\psi}_1$,
\begin{gather}
  D(\mathbf{h}^\dagger) \partial D(\mathbf{h}) \psi = 2^r (m e^\lambda)^h \sqrt{p(z)} \partial \sqrt{p(z)} \psi .
\end{gather}
Note that this ODE is different from the one found in \cite{Dorey:2006an}.
In the next section it will be shown that it is the Langlands dual of $B_{r}^{(1)}$ that gives precisely the ODE associated with the $B_r$ type Lie algebra in their paper.

\subsection{$(B_r^{(1)})^{\vee}=A_{2r-1}^{(2)}$ modified affine Toda equation}

The Langlands dual, obtained by taking the transpose of the Cartan matrix, of $B_r^{(1)}$ is the twisted affine Lie algebra $A_{2r-1}^{(2)}$ for $r \ge 2$.
The $(2r)$-dimensional matrix vector representation with the highest weight $h_1=\omega_1$ is given in appendix \ref{appendix:affine_lie}.
The weight vectors $h_1,\cdots, h_{2r}$ are given by
\begin{equation}
\begin{aligned}
  & h_i-h_{i+1} =\alpha_i \quad (i=1,\cdots, r-1) \,,\\
  & h_{2r+1-i}=-h_{i} \quad (i=1,\cdots r ) \,,\\
  & 2h_r=\alpha_r \,.
\end{aligned}
\end{equation}

After carrying out a gauge transformation \eqref{eq:gauge} with
\begin{gather}  
  U = diag( e^{-\tfrac{\beta}{2} h_1 \cdot \phi},\cdots, e^{-\tfrac{\beta}{2} h_{r} \cdot \phi}, e^{\tfrac{\beta}{2} h_r \cdot \phi},\cdots, e^{\tfrac{\beta}{2} h_{1} \cdot \phi}),
\end{gather}
the linear equation $(\partial + \tilde{A}) \tilde{\Psi} = 0$ gives a set of differential equations.
These differential equations can be combined into a compatibility equation on $\psi = \tilde{\psi}_1$,
\begin{gather}
  D(\mathbf{h}^\dagger) D(\mathbf{h}) \psi = -2^{r-1} (m e^\lambda)^{h} \sqrt{p(z)} \partial \sqrt{p(z)} \psi \label{eq:B2_dual_ODE} .
\end{gather}
This differential equation is the same as the $B$-type equation found in \cite{Dorey:2006an}.

\subsection{$C_r^{(1)}$ modified affine Toda equation}
We discuss the $C^{(1)}_r$ type affine Lie algebra in this section.
Similar to the $B$ type case, we will find that for $C_r^{(1)}$ that we get an ODE that disagrees with \cite{Dorey:2006an}, whereas the $(C_r^{(1)})^\vee$ pseudo-ODE will be in agreement.
The $2r$-dimensional representation with highest weight $h_1=\omega_1$ is given in appendix \ref{appendix:affine_lie}.
The weight vectors denoted $h_1,\cdots, h_{2r}$ are given by
\begin{equation}
\begin{aligned}
  & h_i-h_{i+1}=\alpha_i \quad (i=1,\cdots, r-1) \,,\\
  & h_{2r+1-i}=-h_i \quad (i=1,\cdots, r) \,,\\
  & 2h_r=\alpha_r \,.
\end{aligned}
\end{equation}

After doing a gauge transformation \eqref{eq:gauge} with
\begin{gather}  
  U = diag( e^{-\tfrac{\beta}{2} h_1 \cdot \phi},\cdots, e^{-\tfrac{\beta}{2} h_{r} \cdot \phi}, e^{\tfrac{\beta}{2} h_r \cdot \phi},\cdots, e^{\tfrac{\beta}{2} h_{1} \cdot \phi}),
\end{gather}
the linear equation $(\partial + \tilde{A}) \tilde{\Psi} = 0$ gives a set of differential equations
which can be combined into a compatibility condition on $\psi = \tilde{\psi}_1$,
\begin{gather}
  D(\mathbf{h}^\dagger) D(\mathbf{h}) \psi = (m e^\lambda)^h p(z) \psi .
\end{gather}
These does not coincide with the $C_r$-type pseudo-ODE \cite{Dorey:2006an}, but we will see that the Langlands-dual of $C_r^{(1)}$ gives the correct pseudo-ODE equation.

\subsection{$(C_r^{(1)})^{\vee}=D_{r+1}^{(2)}$ modified affine Toda equation}

The Langlands dual, obtained by taking the transpose of the Cartan matrix, of $C_r^{(1)}$ is the twisted affine Lie algebra $D_{r+1}^{(2)}$ for $r \ge 2$.
The $(2r+2)$-dimensional matrix representation with highest weight $h_1=\omega_1$ is given in appendix \ref{appendix:affine_lie}.
The weight vectors $h_1,\cdots, h_{2r+2}$ of this representations are given by
\begin{equation}
\begin{aligned}
  & h_i-h_{i+1} = \alpha_i \quad (i=1,\cdots, r-1) \,,\\
  & h_{r+1} = h_{r+2} = 0 \,,\\
  & h_{2r+3-i} = -h_i \quad (i=1,\cdots, r) \,.
\end{aligned}
\end{equation}

After doing a gauge transformation \eqref{eq:gauge} with
\begin{gather}  
  U = diag( e^{-\tfrac{\beta}{2} h_1 \cdot \phi},\cdots, e^{-\tfrac{\beta}{2} h_{r} \cdot \phi}, 1, 1, e^{\tfrac{\beta}{2} h_r \cdot \phi},\cdots, e^{\tfrac{\beta}{2} h_{1} \cdot \phi}),
\end{gather}
the linear equation $(\partial + \tilde{A}) \tilde{\Psi} = 0$ gives a set of differential equations 
which can be combined into a compatibility equation on $\psi = \tilde{\psi}_1$,
\begin{gather}
  D(\mathbf{h}^\dagger) \partial D(\mathbf{h}) \psi = 2^{r+1} (m e^\lambda)^{2h} p(z) \partial^{-1} p(z) \psi \label{eq:C2_dual_ODE} .
\end{gather}
This is exactly the same as the $C_r$ type ODE equation of \cite{Dorey:2006an}.

\subsection{$A_{2r}^{(2)}$ modified affine Toda equation}

So far, we have obtained the ODEs associated with classical simple Lie algebras of ABCD type.
New types of (pseudo-)ODEs for $B_r^{(1)}$ and $C_r^{(1)}$ affine Toda equations were found, although the study of these relations to integrable models will be carried out in separate papers.
Here we add another new ODE for $A_{2r}^{(2)}$ to the list of the ODEs.
The remaining twisted classical affine Lie algebra $D_4^{(3)}$ and its exceptional Langlands dual $G_2^{(1)}$ are done in appendix \ref{appendix:exceptional}.

For this $A_{2r}^{(2)}$ case, the $(2r+1)$-dimensional matrix representation with highest weight $h_1=\omega_1$ is given in appendix \ref{appendix:affine_lie}.
The weight vectors $h_1, \ldots, h_r$ are defined by
\begin{equation}
\begin{aligned}
  & h_i - h_{i+1} = \alpha_i \quad (i = 1, \ldots, r-1) \,, \\
  & h_r = \alpha_r \,.
\end{aligned}
\end{equation}
After doing a gauge transformation \eqref{eq:gauge} with
\begin{gather}  
  U = diag( e^{-\tfrac{\beta}{2} h_1 \cdot \phi},\cdots, e^{-\tfrac{\beta}{2} h_{r} \cdot \phi}, 1, e^{\tfrac{\beta}{2} h_r \cdot \phi},\cdots, e^{\tfrac{\beta}{2} h_{1} \cdot \phi}),
\end{gather}
the linear equation $(\partial + \tilde{A}) \tilde{\Psi} = 0$ gives a set of differential equations
which can be combined into a compatibility condition on $\psi = \tilde{\psi}_1$,
\begin{gather}
  D(\mathbf{h}^\dagger) \partial D(\mathbf{h}) \psi = - 2^{r}\sqrt{2} (m e^\lambda)^{h} p(z) \psi .
\end{gather}

The Tzitz\'eica-Bullough-Dodd equation corresponds to the algebra $A_2^{(2)}$.
Since there is only one root, the notation is simplified after setting $\varphi = \omega_1 \cdot \phi$.
Noting that $h_1 = 2\omega_1$, the above analysis shows it should have compatibility condition
\begin{gather}
  (\partial - 2\beta\partial\varphi) \partial (\partial + 2\beta\partial\varphi) \psi = -2\sqrt{2} (m e^\lambda)^3 p(z) \psi .
\end{gather}
This is in agreement with the form of modified Tzitz\'eica-Bullough-Dodd equation obtained in \cite{Dorey:2012bx} after setting $\beta = -1/2$ and $m=1/\sqrt{2}$.

\section{Lie algebra isomorphisms} \label{sec:isomorphisms}

To check the above formalism, it is useful to verify that the standard Lie algebra isomorphisms hold.  This provides a non-trivial check of the (pseudo-)differential equations and the associated linear problems.

\subsection{Rank 1 cases} \label{subsec:Rank1_isomorph}

Many of the matrix representations given in appendix \ref{appendix:affine_lie} break down when $r=1$.  For instance, $D_1^{(1)}$, $(B_1^{(1)})^\vee$, and $C_1^{(1)}$ all reduce to $A_1^{(1)}$ with its 2-dimensional matrix representation.
$B_1^{(1)}$ permits a 3-dimensional matrix representation in a natural way,
and while $(C_1^{(1)})^\vee$ has a 4-dimensional matrix representation, it is actually only rank 3 because the $\psi_3$ component does not come into play, and is equivalent to the $B_1^{(1)}$ case.
This means there are only two cases to consider, $A_1^{(1)}$ and $B_1^{(1)}$.

Letting $e_1$ and $e_2$ be the standard basis vectors for the 2-dimensional $A_1^{(1)}$ case, a connection to the 3-dimensional matrix representation of $B_1^{(1)}$ is found by choosing a basis
  $\{e_1 \otimes e_1$, $\sqrt{2}\,e_1 \otimes e_2$, $e_2 \otimes e_2\}$.
Given a Lie algebra element $x$, it acts on products like $x(e_i \otimes e_j) = x(e_i) \otimes e_j + e_i \otimes x(e_j)$.
This identification gives a perfect match of the matrix representations for the 2- and 3-dimensional cases, and also implies that if $\psi_1$ solves the $A_1^{(1)}$ differential equation $D(-\omega) D(\omega) \psi = (m e^\lambda)^2 p(z) \psi$, then $\psi_1^2$ solves the equation associated with $B_1^{(1)}$, $D(-2\omega) \partial D(2\omega) \psi = 4 (m e^\lambda)^2 \sqrt{p(z)} \partial \sqrt{p(z)} \psi$.

\subsection{$D_2 = A_1 \oplus A_1$} \label{subsec:D2_A1A1_equiv}

In the paper \cite{Dorey:2006an}, the authors showed that this Lie algebra isomorphism was manifest at the level of differential equations.
However for this equivalence to hold, the pseudo-differential equation associated with $D_2$ had to have the $p(z)$ term rescaled.
We will show that this equivalence holds at the level of the linear equation, and furthermore that no re-scaling of $p(z)$ is necessary using the pseudo-differential equations derived here.

For $A_1 \oplus A_1$, the connection breaks into two parts,
\begin{gather}
  A_{A_1 \oplus A_1} = A^{(1)} \oplus A^{(2)} .
\end{gather}
Here, $A^{(i)}$ represents the connection associated with the $i$th node of the Dynkin diagram.
Writing out the connection, it becomes
\begin{gather}
  A^{(i)} = \frac{\beta}{2} \omega_i \cdot \partial \phi \ \sigma^3 + m e^\lambda \left[ e^{\frac{\beta}{2} \alpha_i \cdot \phi} \sigma^+ + p(z) \ e^{-\frac{\beta}{2} \alpha_i \cdot \phi} \ \sigma^- \right] ,
\end{gather}
where $\sigma^3$ and $\sigma^\pm$ are the Pauli matrices.

The matrix representation for algebras of type $D_r$ in appendix \ref{appendix:affine_lie} only hold for $r \ge 3$.
We need a different matrix representation for $D_2$.
The $4$-dimensional representation we use is ($e_{i,j}$ is the matrix with components $(e_{i,j})_{ab} = \delta_{ia} \delta_{jb}$)
\begin{gather}
  E_{\alpha_1} = e_{1,2} + e_{3,4} \,, \quad E_{\alpha_2} = e_{1,3} + e_{2,4}.
\end{gather}
Also note that the weight vectors are $h_1 = \omega_1 + \omega_2$ and $h_2 = -\omega_1 + \omega_2$.
Using the above, the connection becomes
\begin{gather}
  A_{D_2} = \frac{\beta}{2} \begin{pmatrix}
    h_1 &&& \\
    & h_2 && \\
    && -h_2 & \\
    &&& -h_1
  \end{pmatrix} \cdot \partial \phi + m e^\lambda \begin{pmatrix}
    0 & e^{\tfrac{\beta}{2} \alpha_1 \cdot \phi} & e^{\tfrac{\beta}{2} \alpha_2 \cdot \phi} & 0 \\
    p e^{-\tfrac{\beta}{2} \alpha_1 \cdot \phi} & 0 & 0 & e^{\tfrac{\beta}{2} \alpha_2 \cdot \phi} \\
    p e^{-\tfrac{\beta}{2} \alpha_2 \cdot \phi} & 0 & 0 & e^{\tfrac{\beta}{2} \alpha_1 \cdot \phi} \\
    0 & p e^{-\tfrac{\beta}{2} \alpha_2 \cdot \phi} & p e^{-\tfrac{\beta}{2} \alpha_1 \cdot \phi} & 0
  \end{pmatrix} .
\end{gather}
The compatibility equation here is, after an appropriate gauge transformation,
\begin{gather}
  D(-h_1) D(-h_2) \partial^{-1} D(h_2) D(h_1) \ \psi = 4 (m e^\lambda)^2 \sqrt{p(z)} \partial \sqrt{p(z)} \ \psi . \label{eq:D2_comp_eq}
\end{gather}
This agrees with the general form for $D_r^{(1)}$, except the constant in front is $4$, not $2$.

The above connection for $D_2$ can be written using the Kronecker product as
\begin{gather}
  A_{D_2} = \mathbb{1}_2 \otimes A^{(1)} + A^{(2)} \otimes \mathbb{1}_2 .
\end{gather}
So this connection should act on the space of vectors $| v^{(2)} \rangle \otimes | v^{(1)} \rangle$.
The pseudo-differential equation \eqref{eq:D2_comp_eq} acts on the first component of this $4$ component vector which is $\psi^{(1)} \psi^{(2)}$, the product of the upper component corresponding to each of the separate $A_1$ algebras.
Sure enough, $\psi^{(1)} \psi^{(2)}$ can be shown to solve equation \eqref{eq:D2_comp_eq} as expected.

\subsection{$A_3 = D_3$} \label{subsec:A3_D3_equiv}

The two Lie algebras $A_3$ and $D_3$ have the same Dynkin diagram, and represent the same algebra.
By conventions used here, the equivalence of equations of motion is explicit by swapping simple roots $1$ and $2$.
This equivalence can be demonstrated to also hold at the level of the linear problem.
The matrix representation for $A_3$ is $4$-dimensional, while it is $6$-dimensional for $D_3$.
To get an equivalence, it is necessary to map these representations into one another.

Let $\{ e_1, \ldots, e_4 \}$ be the standard basis for the $4$-dimensional space for $A_3$.
Using these basis elements, a $6$-dimensional space can be given by the set of elements $e_{ij} = e_i \wedge e_j$.
To get the correspondence to work exactly, we will need to write out this wedge product explicitly as
  $e_i \wedge e_j = | e_i \rangle_{-1/2} \otimes | e_j \rangle_{1/2} - | e_j \rangle_{-1/2} \otimes | e_i \rangle_{1/2}$,
where the subscript $k = \pm \tfrac{1}{2}$ denotes a Symanzik rotation in the original $A_3$ space.

Given an element $x \in G$ which acts on $e_i$ to give $x(e_i)$, its action on the $6$ dimensional space is $x(e_{ij}) = x(e_i) \wedge e_j + e_i \wedge x(e_j)$.
Ordering the $6$-dimensional basis as $\{ e_{12}, e_{13}, e_{23}, e_{14}, e_{24}, e_{34} \}$ and using the matrix representations in appendix \ref{appendix:affine_lie}, it can be found that the two matrix representations are related through this identification by
\begin{subequations}
\begin{gather}
  A_3(E_{\alpha_1}) = D_3(E_{\alpha_2})  ,\\
  A_3(E_{\alpha_2}) = D_3(E_{\alpha_1})  ,\\
  A_3(E_{\alpha_3}) = D_3(E_{\alpha_3})  ,\\
  A_3(E_{\alpha_0}) = -D_3(E_{\alpha_0}) .\label{eq:A3_D3_matrix_0}
\end{gather}
\end{subequations}
Applying the above Symanzik rotation for $k = \pm \tfrac{1}{2}$ on the $A_3$ linear problem only has the effect of adding a minus sign in front of $p(z)$.
This minus sign cancels out with that of equation \eqref{eq:A3_D3_matrix_0} so that the two linear problems exactly map into one another.

Lastly, the compatibility equations act on the first vector component.
Writing the first component on the $A_3$ side as $| e_1 \rangle_{\pm 1/2} = \psi_{\pm 1/2}$, then the $D_3$ top component is $e_{12} \sim W[\psi_{-1/2}, \psi_{1/2}]$.
So if $\psi$ solves the ODE associated with $A_3$, then $W[\psi_{-1/2}, \psi_{1/2}]$ solves the pseudo-ODE associated with $D_3$, in agreement with \cite{Dorey:2006an}.
Note, however, that like in section \ref{subsec:D2_A1A1_equiv} the factor in front of the $p(z)$ term is just right to get agreement.

\subsection{$B$- and $C$-type isomorphisms}

For the Langlands duals, we have $(B_2^{(1)})^{\vee}=A_{3}^{(2)}$ and $(C_2^{(1)})^{\vee}=D_{3}^{(2)}$.
Since these are twisted versions of $A_3$ and $D_3$, the equivalence between the two follows the same argument as section \ref{subsec:A3_D3_equiv}, with the only change that the basis on the $6$-dimensional side necessary to get agreement is
$\{ e_{12}, \ e_{13}, \ \tfrac{1}{\sqrt{2}}(e_{23}+e_{14}), \ \tfrac{1}{\sqrt{2}}(e_{23}-e_{14}), \ e_{24}, \ e_{34} \}$.
It can also be directly verified that if $\psi^{(1)}$ satisfies equation \eqref{eq:B2_dual_ODE}, then $W[\psi^{(1)}_{-1/2}, \psi^{(1)}_{1/2}]$ satisfies equation \eqref{eq:C2_dual_ODE}.

For the relationship between $B_2^{(1)} = SO(5)$ and $C_2^{(1)} = Sp(4)$, the fundamental matrix representations are $4$-dimensional for $C_2^{(1)}$, and $5$ dimensional for $B_2^{(1)}$.
Luckily, the argument is almost the same as that of section \ref{subsec:A3_D3_equiv} for the $5$-dimensional target basis
$\{ e_{12}, \ e_{13}, \ \tfrac{1}{\sqrt{2}}(e_{23}+e_{14}), \ e_{24}, \ e_{34} \}$
because the element $\tfrac{1}{\sqrt{2}}(e_{23}-e_{14})$ does not play a role.

\begin{table}
\centering
\renewcommand{\arraystretch}{1.5}
\begin{tabular}{|c|c|c|}
  \hline
  $A_r^{(1)}$ & $D(\mathbf{h}) \psi = (-m e^{\lambda})^h p(z) \psi$ & $\mathbf{h} = (h_1, \ldots h_{r+1})$ \\ \hline
  $D_r^{(1)}$ & $D(\mathbf{h}^\dagger) \partial^{-1} D(\mathbf{h}) \psi = 2^{r-1} (m e^\lambda)^h \sqrt{p(z)} \partial \sqrt{p(z)} \psi$ & $\mathbf{h} = (h_1, \ldots h_{r})$ \\ \hline
  $B_r^{(1)}$ & $D(\mathbf{h}^\dagger) \partial D(\mathbf{h}) \psi = 2^r (m e^\lambda)^h \sqrt{p(z)} \partial \sqrt{p(z)} \psi$ & $\mathbf{h} = (h_1, \ldots h_{r})$ \\ \hline
  $A_{2r-1}^{(2)}$ & $D(\mathbf{h}^\dagger) D(\mathbf{h}) \psi = -2^{r-1} (m e^\lambda)^h \sqrt{p(z)} \partial \sqrt{p(z)} \psi$ & $\mathbf{h} = (h_1, \ldots h_{r})$ \\ \hline
  $C_r^{(1)}$ & $D(\mathbf{h}^\dagger) D(\mathbf{h}) \psi = (m e^\lambda)^h p(z) \psi$ & $\mathbf{h} = (h_1, \ldots h_{r})$ \\ \hline
  $D_{r+1}^{(2)}$ & $D(\mathbf{h}^\dagger) \partial D(\mathbf{h}) \psi = 2^{r+1} (m e^\lambda)^{2h} p(z) \partial^{-1} p(z) \psi$ & $\mathbf{h} = (h_1, \ldots h_{r})$ \\ \hline
  $A_{2r}^{(2)}$ & $D(\mathbf{h}^\dagger) \partial D(\mathbf{h}) \psi = -2^{r} \sqrt{2} (m e^\lambda)^{h} p(z) \psi$ & $\mathbf{h} = (h_1, \ldots h_{r})$ \\ \hline
  $G_{2}^{(1)}$ & $D(\mathbf{h}^\dagger) \partial D(\mathbf{h}) \psi = 8 (m e^{\lambda})^{h} \sqrt{p(z)} \partial \sqrt{p(z)} \psi$ & $\mathbf{h} = (h_1, h_2, h_3)$ \\ \hline
  \multirow{4}{*}{$D_{4}^{(3)}$} &  $\bigl[D(\mathbf{h}^\dagger)\partial D(\mathbf{h})+(\zeta+1)2\sqrt{3}(me^\lambda)^4D(\mathbf{h}^\dagger)p(z)$ & \multirow{4}{*}{$\mathbf{h} = (h_1, h_2, h_3)$} \\
                                 & $-\zeta 4\sqrt{3}(me^\lambda)^4 D(-h_1)(\partial p(z)+p(z)\partial) D(h_1)$ & \\
 & $-(\zeta+1)2\sqrt{3}(me^\lambda)^4p(z)D(\mathbf{h})$ & \\
 & $+(\zeta-1)^2 12(me^\lambda)^8 p(z)\partial^{-1}p(z)\bigr]\psi_1=0$ & \\ \hline
\end{tabular}

\caption{Compatibility equations for various algebras, where the weight vectors $h_i$ for each affine algebra are given in sections \ref{sec:A}, \ref{sec:classical}, and appendix \ref{appendix:exceptional}}
\label{tab:odes}
\end{table}

\section{Conclusions and Discussion}

In this paper we have studied the modified affine Toda field equations for affine Lie algebras and their associated linear systems.
By rewriting the linear system in terms of the top component associated with the highest weight, we have obtained the \mbox{(pseudo-)} differential equations which are summarized in Table \ref{tab:odes}.
In the conformal limit, we have shown that these become the (pseudo-)differential equations in the context of the ODE/IM correspondence for classical Lie algebras of ABCD type.
We found that for the affine Lie algebra $\hat{\mathfrak{g}}^{\vee}$, which is the Langlands dual of $\hat{\mathfrak{g}}$, the affine Toda equation leads to the (pseudo-)ODE for the Lie algebra ${\mathfrak{g}}$ of Dorey et al. \cite{Dorey:2006an}.
However, unlike in \cite{Dorey:2006an} where these (pseudo-)differential
equations were conjectured and then verified using various tests, here
we have derived the (pseudo-)differential equations directly from the
linear problem corresponding to the two-dimensional modified affine Toda
equation of $\mathfrak{g}^\vee$.
We note that the matrix form of the ODEs for the classical affine Lie algebras was obtained in \cite{Sun}, which agrees with ours in the conformal limit.

Our choice of form for the Lagrangian \eqref{eq:Lagrangian} gave rise to certain conventions.
In all major equations presented here, $h$ has appeared instead of the more common $h^\vee$.
If one was to use $\alpha_i^\vee$ instead of $\alpha_i$ in equation \eqref{eq:Lagrangian}, then the same results would hold with slight changes.
Everywhere possible dual values would be used, for example $h \leftrightarrow h^\vee$, $\alpha_i \leftrightarrow \alpha_i^\vee$ and $n_i \leftrightarrow n_i^\vee$.
Also, the transpose of the Cartan matrix would now appear in equation \eqref{eq:todaeq2}, which would mean that for non-simply laced cases the choice of matrix representation would have to also switch between dual and non-dual cases.
As a result, if we had instead started with $\alpha_i^\vee$ replacing $\alpha_i$ in \eqref{eq:Lagrangian}, then $\hat{\mathfrak{g}}$ instead of $\hat{\mathfrak{g}}^{\vee}$ would lead to the (pseudo-)ODEs of \cite{Dorey:2006an}.
However, from the point of view of Toda equations our choice of Lagrangian is more natural.

Various isomorphisms between lower rank Lie algebras were also confirmed to hold at both the level of the linear problem, and the (pseudo-)differential equations.
For the cases studied, the factors in front of the $p(z)$ terms for the associated (pseudo-)differential equations were exactly as needed for the agreement to exist.

We also found new ODEs for the affine Lie algebras $B_r^{(1)}$, $C_r^{(1)}$, $A_{2r}^{(2)}$, $D_4^{(3)}$ and $G_2^{(1)}$. 
It would be interesting to explore the integrable systems associated with these algebras.
In particular, for the $B_2$ affine Toda equation, which corresponds to the AdS$_4$ minimal surface, we expect to find the Bethe equations for the homogeneous sine-Gordon model for certain polynomial $p(z)$
  \cite{Hatsuda:2010cc,Hatsuda:2011ke,Hatsuda:2011jn}.
We note that the $A_n^{(1)}$ (non-)affine Toda equations have also appeared in the context of minimal surfaces embedded in the complex projective space $CP_n$ \cite{GriHar,Gervais:1991ds, Dol}.
The ODE/IM correspondence would play an important role for evaluating the area of minimal surfaces embedded in Riemannian symmetric spaces \cite{BolWoo, BoPeWo, Doliwa2, Gervais:1993bf}.

It would also be worthwhile to obtain the ODEs for other exceptional affine Lie algebras and affine Lie superalgebras.
In particular, a supersymmetric generalization would be interesting to explore the integrable structure of superstring theory in AdS space-time.

\section*{Acknowledgments}
We would like to thank J.~Inoguchi, S.~Kobayashi, K.~Sakai, Y.~Satoh and J.~Suzuki
for useful discussions.
The work of K.~I. is supported in part by Grant-in-Aid for Scientific Research from the Japan Ministry of Education, Culture, Sports, Science and Technology. 

\subsection*{Note added:} After submitting this preprint to arXiv, we
learned that Adamopoulou and Dunning obtained the same results for
the $A_r^{(1)}$ case \cite{Adamopoulou:2014fca}.

\appendix
\section{Affine Lie algebras} \label{appendix:affine_lie}

In this appendix we summarize the classical affine Lie algebras and their matrix representations for the fundamental representations.
We explicitly give the matrix representations for $E_\alpha$ where $\alpha$ is a simple root, with the other generators given by $E_{-\alpha} = E_{\alpha}^\top$, and $\alpha^\vee \cdot H = [E_\alpha, E_{-\alpha}]$.  $e_{i,j}$ represents the matrix with components $(e_{i,j})_{ab} = \delta_{ia} \delta_{jb}$.
\begin{description}
\item[$A_r^{(1)}$:]
The highest root is $\theta=\alpha_1+\cdots+\alpha_r$.
The $r+1$-dimensional representation is ($r \ge 1$)
\begin{gather}
  E_{\alpha_0} = e_{r+1,1}, \quad E_{\alpha_i} = e_{i,i+1} .
\end{gather}

\item[$B_r^{(1)}$:]
The highest root is
\begin{align}
  \theta &= \alpha_1 + 2\alpha_2 + \cdots + 2\alpha_{r-1} + 2\alpha_r 
  = \alpha_1^\vee + 2\alpha_2^\vee + \cdots + 2\alpha_{r-1}^\vee + \alpha_r^\vee .
\end{align}
The $2r+1$-dimensional representation is ($r \ge 2$)
\begin{gather}
  E_{\alpha_0} = e_{2r,1} + e_{2r+1,2} , \quad  E_{\alpha_i} = e_{i,i+1} + e_{2r+1-i,2r+2-i} , \notag\\
  E_{\alpha_r} = \sqrt{2} (e_{r,r+1} + e_{r+1,r+2}) .
\end{gather}

\item[$C_r^{(1)}$:]
The highest root is
\begin{gather}
  \theta = 2\alpha_1 + \cdots + 2\alpha_{r-1} + \alpha_r = \alpha_1^\vee + \cdots + \alpha_{r-1}^\vee + \alpha_r^\vee .
\end{gather}
The $2r$-dimensional representation is ($r \ge 1$)
\begin{gather}
  E_{\alpha_0} = e_{2r,1} , \quad E_{\alpha_i} = e_{i,i+1} + e_{2r-i,2r+1-i} , \quad E_{\alpha_r} = e_{r,r+1} .
\end{gather}

\item[$D_r^{(1)}$:]
The highest root is $\theta = \alpha_1 + 2\alpha_2 + \cdots + 2\alpha_{r-2} + \alpha_{r-1} + \alpha_r $.  The $2r$-dimensional representation is ($r \ge 3$)
\begin{gather}
  E_{\alpha_0} = (e_{2r-1,1}+e_{2r,2}), \quad E_{\alpha_i} = (e_{i,i+1}+e_{2r-i,2r+1-i}), \quad E_{\alpha_r} = (e_{r-1,r+1}+e_{r,r+2}) .
\end{gather}

\item[$A_{2r-1}^{(2)} = (B_r^{(1)})^\vee$:]
The highest root is
\begin{align}
  \theta &= \alpha_1^\vee + 2\alpha_2^\vee + \cdots + 2\alpha_{r-1}^\vee + 2\alpha_r^\vee 
  = \alpha_1 + 2\alpha_2 + \cdots + 2\alpha_{r-1} + \alpha_r .
\end{align}
The $2r$-dimensional representation is ($r \ge 2$)
\begin{gather}
  E_{\alpha_0} = e_{2r,2} + e_{2r-1,1} , \quad E_{\alpha_i} = e_{i,i+1} + e_{2r-i,2r+1-i} , \quad E_{\alpha_r} = e_{r,r+1} .
\end{gather}

\item[$A_{2r}^{(2)}$:]
The Coxeter labels are given by (be careful to note that $n_0^\vee = 2$, not $1$)
\begin{gather}
  2 \alpha_0^\vee + 2\alpha_1^\vee + \cdots + 2\alpha_{r-1}^\vee + \alpha_r^\vee = 0 \,,\\
  \alpha_0 + 2\alpha_1 + \cdots + 2\alpha_{r-1} + 2\alpha_r = 0 \,.
\end{gather}
The $(2r+1)$-dimensional matrix representation is given by ($r \ge 1$)
\begin{gather}
  E_{\alpha_0} = e_{2r+1,1} , \quad E_{\alpha_i} = e_{i,i+1} + e_{2r+1-i,2r+2-i} , \quad E_{\alpha_r} = \sqrt{2} (e_{r,r+1} + e_{r+1,r+2}) .
\end{gather}

\item[$D_{r+1}^{(2)} = (C_r^{(1)})^\vee$:]
The highest root is
\begin{gather}
  \theta = 2\alpha_1^\vee + \cdots + 2\alpha_{r-1}^\vee + \alpha_r^\vee = \alpha_1 + \cdots + \alpha_{r-1} + \alpha_r \,.
\end{gather}
The $(2r+2)$-dimensional representation is ($r \ge 2$)
\begin{equation}
\begin{gathered}
  E_{\alpha_0} = \sqrt{2}(e_{r+2,1} + e_{2r+2,r+2}) , \quad   E_{\alpha_i} = e_{i,i+1} + e_{2r+2-i,2r+3-i} , \\
  E_{\alpha_r} = \sqrt{2} (e_{r,r+1} + e_{r+1,r+3}) .
\end{gathered}
\end{equation}

\item[$G_2^{(1)}$:]
The highest root is
\begin{gather}
  \theta = 2\alpha_1 + 3\alpha_2 = 2\alpha_1^\vee + \alpha_2^\vee \,.
\end{gather}
The $7$-dimensional matrix representation is
\begin{equation}
\begin{gathered}
  E_{\alpha_1} = e_{2,3} + e_{5,6} \,,\\
  E_{\alpha_2} = e_{1,2} + e_{6,7} + \sqrt{2} (e_{3,4} + e_{4,5}) \,,\\
  E_{\alpha_0} = e_{6,1} + e_{7,2} \,.
\end{gathered}
\end{equation}

\item[$D_4^{(3)} = (G_2^{(1)})^\vee$:]
The highest root is
\begin{gather}
  \theta = 2\alpha_1^\vee + 3\alpha_2^\vee = 2\alpha_1 + \alpha_2 \,.
\end{gather}
The $8$-dimensional matrix representation is \cite{kac1994infinite}, where $\zeta = \exp(2\pi i/3)$
\begin{equation}
\begin{aligned}
  E_{\alpha_1} &= e_{1,2} + e_{3,4} + e_{3,5} - e_{4,6} - e_{5,6} - e_{7,8} \,,\\
  E_{\alpha_2} &= e_{2,3} - e_{6,7} \,,\\
  E_{\alpha_0} &= e_{4,1} - e_{8,5} + \zeta^2 (e_{6,2} - e_{7,3}) + \zeta (e_{5,1} - e_{8,4}) \,.
\end{aligned}
\end{equation}

\end{description} 

\section{$G_2^{(1)}$ and $D_4^{(3)}$ case} \label{appendix:exceptional}

In this appendix, we will derive an ODE associated with both the exceptional affine Lie algebra $G_2^{(1)}$ and its Langlands dual $D_4^{(3)} = (G_2^{(1)})^\vee$.

For $G_2^{(1)}$, starting with highest weight $\omega_2$ the weight vectors are
\begin{gather}
  h_1 = \omega_2 \,,\\
  h_2 = \omega_1 - \omega_2 \,,\\
  h_3 = 2\omega_2 - \omega_1 \,.
\end{gather}
Writing out the connection using the matrix representation from appendix \ref{appendix:affine_lie} and applying a gauge transformation \eqref{eq:gauge} with 
\begin{gather}
  U = diag( e^{-\tfrac{\beta}{2} h_1 \cdot \phi}, e^{-\tfrac{\beta}{2} h_2 \cdot \phi}, e^{-\tfrac{\beta}{2} h_{3} \cdot \phi}, 1, e^{\tfrac{\beta}{2} h_3 \cdot \phi}, e^{\tfrac{\beta}{2} h_2 \cdot \phi}, e^{\tfrac{\beta}{2} h_{1} \cdot \phi}),
\end{gather}
the component equations can be simply determined.
These equations can be combined into a single differential equation on the top component $\psi = \tilde{\psi}_1$,
\begin{gather}
  D(\mathbf{h}^\dagger) \partial D(\mathbf{h}) \psi = 8 (m e^{\lambda})^{h} \sqrt{p(z)} \partial \sqrt{p(z)} \psi .
\end{gather}

For Langlands dual $D_4^{(3)} = (G_2^{(1)})^\vee$, starting with highest weight $\omega_1$ gives weight vectors
\begin{gather}
  h_1 = \omega_1 \,,\\
  h_2 = \omega_2 - \omega_1 \,,\\
  h_3 = 2\omega_1 - \omega_2 \,.
\end{gather}
Writing out the connection using the matrix representation from appendix \ref{appendix:affine_lie} and applying a gauge transformation \eqref{eq:gauge} with 
\begin{gather}
  U = diag( e^{-\tfrac{\beta}{2} h_1 \cdot \phi}, e^{-\tfrac{\beta}{2} h_2 \cdot \phi}, e^{-\tfrac{\beta}{2} h_{3} \cdot \phi}, 1, 1, e^{\tfrac{\beta}{2} h_3 \cdot \phi}, e^{\tfrac{\beta}{2} h_2 \cdot \phi}, e^{\tfrac{\beta}{2} h_{1} \cdot \phi}),
\end{gather}
the component equations can be determined.
Due to $E_{\alpha_0}$ having so many components, the linear problem has many $p(z)$ terms, which makes it difficult to write out a simple equation in terms of just $\psi = \tilde{\psi}_1$.
However, it can be done, giving
\begin{align}
&\Bigl[
D(\mathbf{h}^\dagger)\partial D(\mathbf{h})
+(\zeta+1)2\sqrt{3}(me^\lambda)^4D(\mathbf{h}^\dagger)p(z)-\zeta 4\sqrt{3}(me^\lambda)^4 D(-h_1)(\partial p(z)+p(z)\partial) D(h_1)
\nonumber\\
&
-(\zeta+1)2\sqrt{3}(me^\lambda)^4p(z)D(\mathbf{h})
+(\zeta-1)^2 12(me^\lambda)^8 p(z)\partial^{-1}p(z)
\Bigr]\psi_1=0
\label{eq:D_4-except_eq}
\end{align}

\end{document}